# Ferro-type order of magneto-electric quadrupoles as an order-parameter for the pseudo-gap phase of a cuprate superconductor

S. W. Lovesey[1,2], D. D. Khalyavin[1] and U. Staub[3]

[1]ISIS Facility, STFC Oxfordshire OX11 0QX, UK,

[2]Diamond Light Source Ltd, Oxfordshire OX11 0DE, UK,

[3]Swiss Light Source, Paul Scherrer Institut, CH 5232 Villigen PSI, Switzerland

Understanding the pseudo-gap phase from which a superconducting state emerges is a central problem in the physics of cuprates. The phase exists below a temperature $T^*$ for less than optimal doping of holes, and no clear jump in the specific heat or change in lattice symmetry is observed on crossing $T^*$. Here we show that, two sets of intriguing experimental results in the literature concerning magnetic properties of underdoped YBCO are united in a single theory based on Cu quadrupole moments. The quadrupole is a product of spin and the electric dipole, so there is no magnetic dipole moment, and standard crystallography describes the magnetic motif. One experiment employed the Kerr effect and the second Bragg diffraction of polarized neutrons, and both are shown to imply an order-parameter for the pseudo-gap phase composed of magneto-electric quadrupoles at in-plane Cu sites in a ferro-type order.

The successful interpretation of neutron diffraction data for YBCO yields the first experimental evidence that neutrons are deflected by magneto-electric multipoles. Predicted in 2014, our verification of the effect validates a new application for the experimental technique of choice in studies of magnetic motifs and excitations.

Mechanisms for superconductivity in ceramic materials formed from layers paved with squared $CuO_2$ plaquettes remain unsettled, almost three decades after the surprising property was discovered in a barium-doped compound of lanthanum and copper oxide[1]. A general phase diagram of cuprates starts with an antiferromagnetic parent compound that transforms when doped to a regime in which superconductivity appears (underdoped regime), reaching an optimally doped region with a maximum temperature for superconductivity. A pseudo-gap phase appears below a temperature $T^*$ in underdoped materials, which is characterized by a wave-vector dependent opening of a (pseudo) gap at the Fermi surface[2]. However, there is no clear structural modification or a distinct jump in specific heat measurements associated with $T^*$.

$YBa_2Cu_3O_{6+x}$ (YBCO) has superconducting transition temperatures above that of boiling liquid nitrogen. It crystallizes in a layered-type structure with cell dimensions a ≈ 3.85 Å, c ≈ 11.7 Å. A unit cell (Fig. 1) contains three pseudo-cubic elementary perovskite unit cells and there are two $CuO_2$ plaquettes in a unit cell approximately 3.2 Å apart. The relation between oxygen concentration x and hole doping is non-trivial in YBCO, because the unit-cell contains two Cu types. Electronic properties are almost 2-dimensional and display strong angular anisotropy[2]. A gap associated with an insulating phase only exists for electrons travelling parallel to Cu-O bonds, whilst electrons travelling at 45° to this bond can move freely throughout the crystal. Low energy electrons reside in a single band of carriers formed by hybridization of $Cu^{2+}$ orbitals and oxygen 2p-electrons. An ordering wavevector consistent with a doubling of the in-plane chemical unit-cell describes the antiferromagnetic order in the $CuO_2$ planes of the parent compound. This order disappears very fast when the material is doped.

Compelling evidence for long-range magnetic order in the pseudo-gap phase in the underdoped region of the phase diagram has come from elegant Bragg diffraction experiments on crystals of YBCO with different hole doping, using the technique of neutron polarization analysis[3, 4, 5]. Key findings from the studies are that magnetic order in the pseudo-gap phase of YBCO is indexed on the chemical structure, and the onset temperature matches approximately $T^*$. No less compelling evidence for magnetic order is observation of the rotation of polarization of reflected light (Kerr effect) that is a direct manifestation of broken time-reversal symmetry generally associated with long-range order in which there is a net magnetic moment[6, 7, 8]. Both sets of observation, neutron Bragg diffraction and Kerr effect, are fully explained by us using a ferro-type order of magneto-electric quadrupoles depicted in Fig. 2. To achieve this new direction in understanding the physical properties of cuprate superconductors it was necessary to prove that neutrons, like photons, are diffracted by magneto-electric multipoles that are both time-odd and parity-odd[9].

We begin an account of our findings with steps in the interpretation of Bragg diffraction data, because it requires knowledge of the full magnetic space-group with all translations, rotations and improper-rotations. On the other hand, the corresponding magnetic crystal-class (point group) delineates bulk properties including the Kerr effect.

In the confines of a conventional analysis of magnetic neutron diffraction some Bragg spots from underdoped YBCO appearing at $T^*$ can be attributed to magnetic dipole moments ≈ 0.1 $\mu_B$ inclined at about 45º to the c-axis[4]. The dipole moment is revised down to ≈ 0.05 $\mu_B$ for nearly optimally doped YBCO while the tilt is similar in both samples[5]. Because magnetic neutron diffraction is made only by dipole moments normal to the Bragg wavevector, no discernible intensity in the (0, 0, 2) Bragg spot implies that magnetic dipole moments in the a-b plane are absent and the dipole moment is actually confined to the c-axis[4, 5].

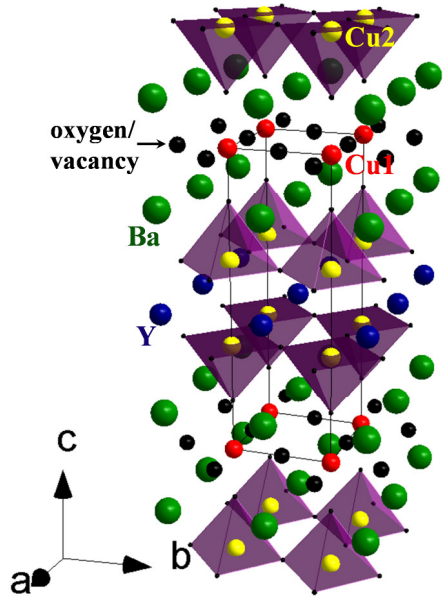

Figure 1. Crystal structure of YBaCu$_3$O$_7$ with the magnetic Cu2 ions in the CuO$_2$ planes.

To reconcile the apparently conflicting information from a conventional analysis we started from the parent structure P4/mmm-type with Cu2 ions using sites 2g (symmetry 4mmm). We derived magnetic space-groups with broken time-reversal symmetry and three specified properties: bulk ferromagnetism is forbidden, an antiferromagnetic motif of magnetic moments parallel to the c-axis is allowed, and anapoles (toroidal dipole moments) in the a-b plane are allowed. Refining the outcome of this exercise against Bragg diffraction data leads to a unique answer for the space group that turns out to possess more symmetry than we specified, namely, Cm'm'm' with Cu2 in sites 4k with symmetry m'm'2 that does not include a centre of inversion symmetry. Our proposal for the magnetic space-group can be explored further in future experiments.

For Cm'm'm' the Bragg spot (0, 0, 2) has no intensity, as required by the measurements, and the magnetic dipole moment $\langle(\mathbf{L} + 2\mathbf{S})_z\rangle$ and magneto-electric quadrupole of Cu2 ions are permitted sources of diffraction at the Bragg spot (1, 0, 1) at which polarization analysis was applied[4]. The corresponding magnetic crystal-class m'm'm' is neither polar nor chiral. A notable feature of the crystal class m'm'm' is that it allows the Kerr effect[6, 7, 8].

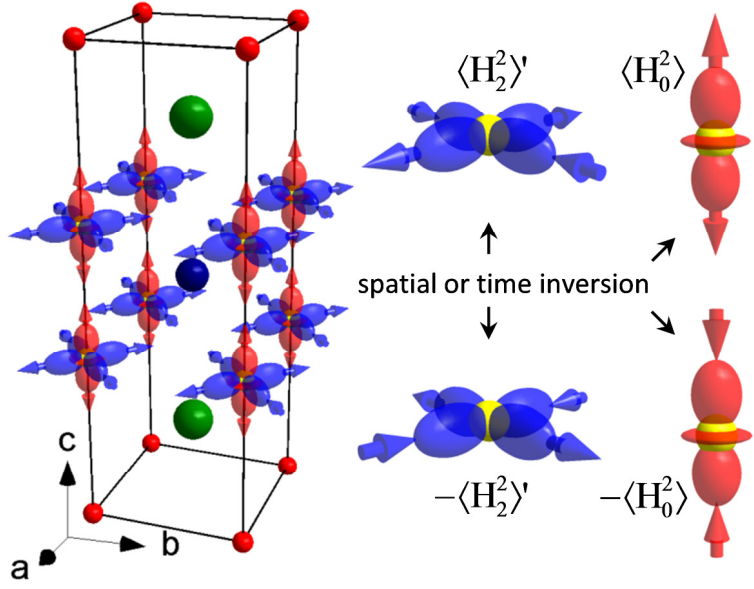

Figure 2. Ferro-type ordering of magneto-electric quadrupoles in CuO$_2$ planes for YBCO in the pseudo-gap phase. Arrows indicate spin directions in magnetic charge-like quadrupoles $\langle H^2_0 \rangle \propto \langle 3S_z n_z - \mathbf{S} \cdot \mathbf{n} \rangle$ and $\langle H^2_{+2} \rangle' \propto \langle S_x n_x - S_y n_y \rangle$, together with their response to spatial or time inversion.

The amplitude for magnetic neutron Bragg diffraction with a scattering wavevector **k** ≡ (h, k, l) and Miller indices h, k, l can be derived from[10],

$$\langle \mathbf{Q} \rangle = \langle \exp(i\mathbf{R} \cdot \mathbf{k})\, [\mathbf{S} - i(\mathbf{k} \times \mathbf{p})/\hbar k^2] \rangle, \qquad (1)$$

where **R**, **S** and **p** are electron position, spin and linear momentum operators, respectively. The scattering amplitude is $\langle \mathbf{Q}_\perp \rangle = [\boldsymbol{\kappa} \times (\langle \mathbf{Q} \rangle \times \boldsymbol{\kappa})]$ using a unit vector $\boldsymbol{\kappa} = \mathbf{k}/k$. In these expressions, angular brackets ⟨ ... ⟩ denote the expectation value, or time-average, of the enclosed operator. For Cu2 ions using the magnetic space-group Cm'm'm' and a reflection (1, 0, 1) we find,

$$(\langle Q_y \rangle - \langle Q_x \rangle) \approx -12i\sqrt{(1/5)}\cos(\varphi)\,\kappa_x \langle H^2_{+2} \rangle',$$

$$(\langle Q_y \rangle + \langle Q_x \rangle) \approx -2i\sqrt{(6/5)}\cos(\varphi)\,\kappa_x \langle H^2_0 \rangle,$$

$$\langle Q_z \rangle \approx 3i\sin(\varphi)\,\langle (L + 2S)_z \rangle + 2i\sqrt{(6/5)}\cos(\varphi)\,\kappa_z \langle H^2_0 \rangle. \qquad (2)$$

In (2) we omit high-order multipoles that are most likely very small compared to the retained dipole and quadrupoles[9]. Orthogonal Cartesian (x, y, z) match {(1, −1, 0), (1, 1, 0), (0, 0, 1)}, and the phase $\varphi = 2\pi l z$ with position parameter $z \approx 0.36$. Magneto-electric quadrupoles in (2) are $\langle H^2_0 \rangle \propto \langle 3S_z n_z - \mathbf{S} \cdot \mathbf{n} \rangle$ and $\langle H^2_{+2} \rangle' \propto \langle S_x n_x - S_y n_y \rangle$ where **n** is the electric dipole operator. Magneto-electric multipoles of any rank are strictly forbidden if the ion in question occupies a site that is a centre of inversion symmetry, but this is not the case for the space group we propose. The solution (2) does not require $\langle (L + 2S)_z \rangle$ to be non-zero to obtain $\langle Q_z \rangle \neq 0$ and naturally resolves the aforementioned conundrum in a conventional analysis. With $\langle (L + 2S)_z \rangle = 0$ the dipole field is zero at a great distance from a Cu ion.

Experimental data for spin-flip (SF) intensities are displayed in Fig. 3. Corresponding expressions can be derived from (2) using SF = $|\langle \mathbf{Q}_\perp \rangle|^2 - |\mathbf{P} \cdot \langle \mathbf{Q}_\perp \rangle|^2$, with primary polarization of the neutron beam **P** a unit vector. Previously we mentioned the measurement SF = 0 at (0, 0, 2) with **P** // **k**. For reflection (1, 0, 1) one has **k**/k = $(\kappa_a, 0, \kappa_c)$ with $(\kappa_a^2 + \kappa_c^2)$ = 1 and $\kappa_a \approx 0.950$. Three configurations of **P** and **k** labelled (a), (b) and (c) in Fig. 3 were investigated and the corresponding expressions for SF are simple to find,

(a) **P** // **k**, SF = $|\langle Q_\perp \rangle|^2$.

(b) **P** • **k** = 0 using **P** = $(-\kappa_c, 0, \kappa_a)$, and **P** • $\langle Q_\perp \rangle$ = $(\kappa_a \langle Q_c \rangle - \kappa_c \langle Q_a \rangle)$ leads to SF = $|\langle Q_b \rangle|^2$.

(c) **P** • **k** = 0 using **P** = (0, 1, 0), and **P** • $\langle Q_\perp \rangle$ = $\langle Q_b \rangle$ leads to SF = $|\kappa_a \langle Q_c \rangle - \kappa_c \langle Q_a \rangle|^2$.

Data sets in Fig. 3 are not independent since SF(a) = SF(b) + SF(c)[4].

Setting $\langle (L + 2S)_z \rangle = 0$ in (2) leads to a ratio of spin-flip intensities that we have drawn in Fig. 3,

$$\text{SF(c)/SF(b)} = (\langle H^2_0 \rangle / \langle H^2_{+2} \rangle')^2 \, (1/6) \, (\kappa_c + \kappa_a (a/c)\sqrt{2})^2. \qquad (3)$$

Experimental data gives SF(c)/SF(b) ≈ 0.9 ± 0.2 and yields $(\langle H^2_0 \rangle / \langle H^2_{+2} \rangle')^2 \approx 9.5 \pm 2$ that is independent of temperature in the pseudo-gap phase, to a good approximation.

Note that there is no direct relation between our theory, using magnetic charge at Cu2 sites, and loops of current in $CuO_2$ planes[11, 12]. Such loops create a magnetic dipole moment from orbital angular momentum that is off-set from Cu sites, and this construct alone does not explain the neutron diffraction under consideration[4, 5].

Magneto-electric multipoles merit more discussion in view of the fact that they are central in our theory for neutron diffraction by YBCO. Treating **k** in (1) as a small quantity $\langle Q \rangle$ contains a spin moment $\langle S \rangle$ at the initial level of an expansion. Some algebra is required to show that, **R** and **p** in (1) produce a contribution proportional to orbital angular momentum, **L** = **R** x **p**, in the second level of an expansion[10, 15]. Combining the two results, we may write $\langle Q \rangle \approx (1/2)$ f(k) $\langle L + 2S \rangle$ in which f(k) is a normalized atomic form factor, and f(0) = 1 is used in (2) to keep notation simple. This well-established, small **k** limit for $\langle Q \rangle$ is routinely exploited to determine size and motif of ordered magnetic dipoles. At the second level in an expansion in **k**, the spin contribution to $\langle Q \rangle$ is parity-odd, namely, $ikR\langle(\kappa \cdot n)S\rangle$ in which **n** = **R**/R. For this contribution we can write,

$$\langle(\kappa \cdot n)S\rangle \equiv (1/2)[\kappa \times \langle S \times n \rangle + \langle S(\kappa \cdot n) + (\kappa \cdot S)n - (2/3)\kappa(S \cdot n)\rangle].$$

The spin anapole $\langle S \times n \rangle$ has been observed using resonant x-ray diffraction, e.g., refs. 16, 17. The quadrupole in (2) arises from $\langle S(\kappa \cdot n) + (\kappa \cdot S)n - (2/3)\kappa(S \cdot n)\rangle$. A magnetic charge $\langle S \cdot n \rangle$ is omitted from the development of $\langle(\kappa \cdot n)S\rangle$, because it is multiplied by $\kappa$ and does not therefore appear in the scattering amplitude $\langle Q_\perp \rangle$. (The label magnetic charge is justified by the observation that a "magnetic charge" inserted in Maxwell's equations, with symmetries of the electric and magnetic field unchanged, is both time-odd and parity-odd.)

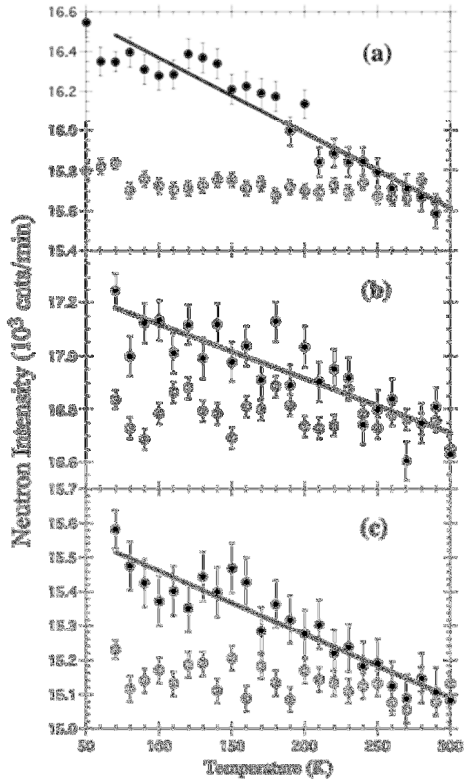

Figure 3. Spin-flip (SF) intensities of the Bragg spot **k** = (1, 0, 1) from YBCO as a function of temperature are shown as solid circles. Open circles indicate polarization leakage that is significantly less than the SF data of interest. Equation (3) for the ratio of intensities is drawn as a solid line. In panel (a) neutron polarization **P** and wavevector **k** are parallel. On the other hand, **P** and **k** are orthogonal in panels (b) and (c), with **P** parallel to the b-axis in (b) and **P** in the a-c plane in (c). Polarization analysis is employed in diffraction to separate magnetic from nuclear signals that are superimposed because the magnetic motif coincides with chemical structure. Data are reproduced from Fig. 5 of ref. 4.

The lack of inversion symmetry at sites 4k in Cm'm'm' used by Cu2 ions allows significant admixtures of 2p(O) and 3d(Cu) orbitals involved in bonding and 3d(Cu) and 4p(Cu) orbitals, states with different orbital angular momentum. In a recent theoretical study the quantum model of a magnetic charge coupled to a quantum rotator introduced by Stone has been adapted to incorporate a crystal field, and it describes magneto-electric quadrupoles appearing in (2) and (3)[13, 14]. Magneto-electric multipoles have not only been used recently to interpret resonant x-ray diffraction and absorption data[16, 17], but also have been found in DFT simulations of magneto-electric materials[18]. The atomic form factor associated with magneto-electric quadrupoles in YBCO has been calculated for 3d(Cu) and 4p(Cu) orbitals using output from a code for atomic wavefunctions. The maximum in the calculated form factor occurs at k ≈ 1.7 Å$^{-1}$ that almost perfectly matches the magnitude of the (1, 0, 1) wavevector analysed by polarization analysis, a finding which adds confidence to our interpretation.

In summary, a measured Kerr effect[6, 7, 8] and magnetic Bragg spots[3, 4, 5] observed with underdoped YBCO are unified in a single theory that exploits parity-odd magnetic multipoles in a ferro-type order. It is satisfying that the far-field of a magnetic Cu ion need not contain a dipole, and the necessary concepts and methods are well-established. Moreover, an observation of anapoles (toroidal dipoles) in the parent compound CuO suggests the role of magneto-electric multipoles should not be altogether surprising[9, 17]. Available measurements on YBCO infer that magneto-electric quadrupoles localized on Cu2 ions can act as the order-parameter in the pseudo-gap phase. For Bragg diffraction, our theory overcomes conflicting information deduced from a conventional interpretation of measurements that is based on magnetic (parity-even) dipoles alone.

We have verified the prediction that neutrons are deflected by magneto-electric multipoles[9]. The effect is allowed for magnetic ions that occupy sites that are not centres of inversion symmetry.

**Acknowledgements.** We thank Dr. P. Bourges for comments on an early draft of the communication and making data for Fig. 3 available to us. Professor G. van der Laan performed the calculation of the quadrupole form factor that we refer to. We have benefited from discussions with Professor B. Keimer, Dr. C. Niedermayer and Dr. M. Ramakrishnan. This work was in part supported by the Swiss National Science Foundation.